\documentstyle[12pt]{article}\textheight 230mm\textwidth 150mm
\newcommand{\be}{\begin{eqnarray}}
\newcommand{\ee}{\end{eqnarray}}
\newcommand{\eel}[1]{\label{#1}\end{eqnarray}}
\newcommand{\r}[1]{(\ref{e:#1})}

\newcommand{\vb}{{\cal h}}
\newcommand{\hb}{{\cal i}}
\newcommand{\lra}{{\leftrightarrow}}

\newcommand{\ra}{{\rightarrow}}
\newcommand{\nn}{\nonumber}
\newcommand{\eg}{{\em e.g.\ }}
\newcommand{\ie}{{\em i.e.\ }}

\newcommand{\al}{\alpha}
\newcommand{\ga}{{\gamma}}
\newcommand{\la}{{\lambda}}

\newcommand{\del}{{\delta}}

\newcommand{\pet}{{\cal P}}

\newcommand{\bata}{\bar{\eta}}
\newcommand{\bapet}{\bar{\pet}}

\newcommand{\bac}{\bar{c}}

\newcommand{\bak}{\bar{k}}

\newcommand{\bett}{{\bf 1}}
\newcommand{\halv}{\frac{1}{2}}

\begin{document}
\begin{titlepage}
\noindent
G\"{o}teborg ITP 93-17\\
May 1993\\

\vspace*{10 mm}
\begin{center}{\LARGE\bf Gauge fixing and abelianization\\ in simple BRST
quantization.}\end{center}
\vspace*{3 mm} \begin{center} \vspace*{12 mm}

\begin{center}Robert Marnelius \\
\vspace*{7 mm}
{\sl Institute of Theoretical Physics\\
Chalmers University of Technology\\
S-412 96  G\"{o}teborg, Sweden}\end{center}
\vspace*{15 mm}
\begin{abstract}
In a previous paper \cite{Simple} it was shown that the BRST charge $Q$ for any
gauge model with a
Lie algebra symmetry may be decomposed as
$Q=\del+\del^{\dag},\;\;\;\del^2=\del^{\dag 2}=0,\;\;\;[\del,
\del^{\dag}]_+=0$ provided dynamical Lagrange multipliers are used but without
introducing other matter variables in $\del$ than the gauge generators in $Q$.
In this paper further decompositions are derived but now by means of gauge
fixing operators.
As in \cite{Simple} it is shown that $\del=c^{\dag a}\phi_a$ where $c^a$ are
new ghosts and $\phi_a$
are nonhermitian variables satisfying the gauge algebra. However, in
distinction to \cite{Simple} also
solutions of the form $\del=c^{\dag a}A_a$ where $A_a$ satisfy an abelian
algebra is derived
(abelianization).  By
means of a bigrading the BRST condition reduces to
$\del|ph\hb=\del^{\dag}|ph\hb=0$ on inner product spaces whose general
solutions are expressed in
terms of the solutions to a proper Dirac quantization. Thus, the procedure
provides for inner
products for the solutions of a Dirac quantization.
\end{abstract}\end{titlepage}

\setcounter{page}{1}
\section{Introduction.}
In a previous paper \cite{Simple} the BRST quantization on inner product spaces
was considered for
gauge theories with a general Lie group gauge invariance.  It
was shown that provided dynamical Lagrange multipliers are introduced the BRST
charge $Q$ may
always be written in the form
\be
&&Q=\del+\del^{\dag}
\eel{e:1}
 where
\be
&&\del^2=\del^{\dag 2}=0
\eel{e:2}
and
\be
&&[\del, \del^{\dag}]_+=0
\eel{e:3}
consistent with a nilpotent $Q$. These properties are
 necessary conditions \cite{HM,RM} for  a quantization on an inner product
space. By means of a
bigrading of the original state space the BRST condition
\be
&&Q|ph\hb=0
\eel{e:4}
may be replaced by
\be
&&\del|ph\hb=\del^{\dag}|ph\hb=0
\eel{e:5}
which was solved explicitly.  The solutions were shown to have a very simple
form which is the
reason why it was called simple BRST quantization. In this paper we will
analyze this approach
further. We shall then find new decompositions \r{1} leading to solutions of
\r{5} in new forms.

\section{Decompositions of general BRST charges.}
\setcounter{equation}{0}
As in \cite{Simple} we consider for simplicity a general gauge model with
finite number
of degrees of freedom in which the gauge group is a Lie group. Within the
Hamiltonian formulation of the corresponding BRST invariant model the BRST
charge may be
chosen to be in the general BFV form \cite{BV} ($a, b, c =1,\ldots,m<\infty$)
\be
&&Q=\psi_a\eta^a-\frac{1}{2}iU_{bc}^{\;\;a}\pet_a
\eta^b\eta^c-\frac{1}{2}iU_{ab}^{\;\;b}\eta^a + \bapet_a\pi^a
\eel{e:201}
where $\psi_a$ are the bosonic gauge generators
(constraints) satisfying
\be
&[\psi_a, \psi_b]_{-}=iU_{ab}^{\;\;c}\psi_c
\eel{e:202}
where $U_{ab}^{\;\;c}$ are the structure constants.
 $\eta^a$ and $\bata^a$ are Faddeev-Popov (FP)
ghosts and antighosts respectively, and $\pet_a$ and $\bapet_a$
their conjugate momenta. $\pi_a$ is the conjugate momentum to the Lagrange
multiplier $v^a$.
Their fundamental algebra is (the nonzero part)
 \be
&[\eta^a, \pet_b]_+=[\bata^a, \bapet_b]_+=\del^a_{\;b},\;\;\;[\pi_b,
v^a]_-=-i\del^a_{\;b}
\eel{e:203}
which when combined with \r{202} makes $Q$ nilpotent. Since $Q$ is required to
have ghost number one,
$\eta^a$ and $\bapet_a$ have ghost number  one while $\bata^a$
and $\pet_a$ have ghost number minus one. The ghost number operator is
\be
&&N=\eta^a\pet_a-\bata_a\bapet^a
\eel{e:204}
Since $Q$ is required to be hermitian, $\psi_a$ must either be hermitian or
contain hermitian
conjugate pairs. Here we shall choose $\psi_a$ as well as all
the other variables to be hermitian which is in fact a choice one always may
do.

As a first step to decompose \r{201} according to \r{1} we introduce
the following non-hermitian  ghosts
\be
&&c^a\equiv\halv(\eta'^a-i\bapet'^a)\nn\\
&&k_a\equiv\pet'_a-i\bata'_a
\eel{e:205}
where the ghosts $\eta'_a, \pet'_a, \bata'_a$ and $\bapet'_a$ are obtained from
$\eta_a, \pet_a,
\bata_a$ and $\bapet_a$ by a unitary transformation. From \r{203} this implies
that they satisfy
 the algebra (the nonzero part)
\be
&&[k_a, c^{\dag b}]_+=\del^b_{\;a}
\eel{e:206}
If we require  this unitary transformation also to preserve ghost number, then
$c^a$ has ghost number
one while $k_a$ has ghost number minus one. The ghost number operator is then
\be
&&N=\eta'^a\pet'_a-\bata'_a\bapet'^a=c^{\dag a}k_a-k^{\dag }_ac^a
\eel{e:207}

We may now define $\del$ by
\be
&&\del=[c^{\dag a}k_a, Q]
\eel{e:208}
which automatically imply the decomposition \r{1}.
However, this   $\del$ is {\em
not} nilpotent in general.
The origin of this problem is seen when we consider a decomposition according
to the bigrading
generated by $c^{\dag a}k_a$ and $k^{\dag }_ac^a$
\be
&&Q=\sum_{n,m}Q_{n,m}
\eel{e:209}
where
\be
&&[c^{\dag a}k_a, Q_{n,m}]=nQ_{n,m} \nn\\
&&[k^{\dag }_ac^a, Q_{n,m}]=-mQ_{n,m}
\eel{e:210}
The reason is that $\del$ in \r{208} can only be nilpotent if
\be
&&[k^{\dag }_ac^a, \del]=0
\eel{e:211}
which requires $\del=Q_{1,0}$ and $\del^{\dag}=Q_{0,1}$. Hence, the unitary
transformation above
must reduce the decomposition \r{209} according to the following form
\be
&&Q=Q_{0,1}+Q_{1,0}
\eel{e:212}
which is just an alternative form of \r{1}. As an example we may consider the
identity
transformation such that
\be
&&c^a\equiv\halv(\eta^a-i\bapet^a)\nn\\
&&k_a\equiv\pet_a-i\bata_a
\eel{e:213}
In this case the BRST charge \r{201} is
\be
&&Q=Q_{0,1}+Q_{1,0}+Q_{2,1}+Q_{1,2}
\eel{e:214}
where
\be
&&Q_{2,1}=-\frac{1}{4}iU_{bc}^{\;\;a}k^{\dag}_a
c^{\dag b}c^{\dag c}\nn\\
&&Q_{2,1}=-\frac{1}{4}iU_{bc}^{\;\;a}k_a
c^b c^c
\eel{e:215}
Thus, in this case the decomposition \r{1} satisfying \r{2} is only obtained
for an abelian model.
In the general case we need a non-trivial unitary transformation in order  to
transform away the
terms  \r{215} in $Q$. The general form of such a unitary transformation  is
\be
&&\eta'^a=A^a_{\;\;b}\eta^b + B^a_{\;\;b}\bapet^b,\;\;\;
\pet'_a=C^b_{\;\;a}\pet_b + D^b_{\;\;a}\bata_b\nn\\
&&\bapet'^a=E^a_{\;\;b}\bapet^b + F^a_{\;\;b}\eta^b,\;\;\;
\bata'_a=G^b_{\;\;a}\bata_b + H^b_{\;\;a}\pet_b
\eel{e:216}
together with the implied transformations of the other variables. The operator
valued matrices
$A,\dots,H$ must have ghost number zero  and must satisfy the conditions which
makes \r{216} an
invertible unitary transformation. In \cite{Simple} it was assumed that these
matrices
only depend on
the Lagrange multipliers $v^a$ and it was then found that the following unitary
transformation
transformed away terms of the form \r{215}
\be
&&\eta'^a=\eta^a, \;\;\;
\pet'_a=\pet_a + (M^{-1})^ b_{\;\;c}K^c_{\;\;a}\bata_b\nn\\
&&\bapet'^a=(M^{-1})^a_{\;\;b}\bapet^b -
(M^{-1})^a_{\;\;b}K^b_{\;\;c}\eta^c\nn\\
&&\bata'_a=M^b_{\;\;a}\bata_b
\eel{e:217}
 for one specific choice of the
matrices $M$ and $K$. For this transformation $Q$ was shown to decompose
according to
\r{1}. Furthermore, it was found that
 \be &&\del=\phi'_ac^{\dag a}=c^{\dag a}\phi_a
\eel{e:218}
where in turn the non-hermitian expressions $\phi^{(')}_a$ were shown to
satisfy
\be
&&[\phi^{(')}_a, \phi^{(')}_b]=iU_{ab}^{\;\;c}\phi^{(')}_c
\eel{e:219}

In the next section another unitary transformation is given which also leads to
the decomposition
\r{1} with similar properties as the above one.

\section{Decompositions by means of gauge fixing.}
\setcounter{equation}{0}

Consider a set of hermitian gauge fixing operators $\chi^a$ for which the
matrix operator
\be
&&M^a_{\;\;b}\equiv i[\chi^a, \psi_b]
\eel{e:301}
is assumed to have an inverse. In analogy with the transformation \r{217}
obtained in \cite{Simple}
we try a transformation of the form
\be
&&\bata'_a=M^b_{\;\;a}\bata_b
\eel{e:302}
which together with
\be
&&\bapet'^a=(M^{-1})^a_{\;\;b}\bapet^b,
\;\;\;\eta'^a=\eta^a,\;\;\;\pet'_a=\pet_a
\eel{e:303}
constitute a unitary transformation provided the different components of
$M^a_{\;\;b}$ commute which
we assume. We then define a gauge generator which commutes with the new ghosts
\be
&&\psi'_a\equiv\psi_a+\halv(\bata_b\bapet^c-\bapet^c\bata_b)G^b_{\;\;ca}\nn\\
&&=\psi_a+\halv(\bata'_b\bapet'^c-\bapet'^c\bata'_b)H^b_{\;\;ca}
\eel{e:304}
where
\be
&&G^b_{\;\;ca}\equiv(M^{-1})^d_{\;\;c}[M^b_{\;\;d}, \psi_a]\nn\\
&&H^b_{\;\;ca}\equiv(M^{-1})^b_{\;\;d}[M^d_{\;\;c}, \psi_a]
\eel{e:305}
where we assume that $[M^b_{\;\;d}, \psi_a]$ commutes with $M^f_{\;\;c}$. Under
this
assumption we also have
\be
&&[\psi'_a, \psi'_b]=iU_{ab}^{\;\;c}\psi'_c
\eel{e:306}
The BRST charge \r{201} may now be written as \r{1} with
\be
&&\del=\phi'_ac^{\dag a}=c^{\dag a}\phi_a
\eel{e:307}
where
\be
&&\phi_a=\psi'_a+\halv
H^b_{\;\;ab}-i\pi_bM^b_{\;\;a}-H^b_{\;\;ca}k^{\dag}_bc^c+\halv
iU_{ca}^{\;\;b}c^{\dag c}k_b\nn\\
&&\phi'_a=\phi_a-\halv iU_{ab}^{\;\;b}
\eel{e:308}
where $H^b_{\;\;ca}$ is defined in \r{305} and where $c^a$ and $k_a$ are
defined by \r{205}. Notice,
however, that $\phi_a$ and $\phi'_a$ in \r{308} are ambiguously defined. We may
make the replacement
\be
&&\phi_a\ra\phi_a+c^{\dag b}k_cf_{ab}^{\;\;\;c}\nn\\
&&\phi'_a\ra\phi'_a+g_{ab}^{\;\;\;c}k_cc^{\dag b}
\eel{e:309}
without affecting $\del$ provided $f_{ab}^{\;\;\;c}$ and $g_{ab}^{\;\;\;c}$ are
symmetric in the
indices $a$ and $b$ which is obvious from \r{307}. This freedom we may use to
make $\phi_a$ and
$\phi'_a$ in \r{308} satisfy the algebra \be
&&[\phi^{(')}_a, \phi^{(')}_b]=iU_{ab}^{\;\;c}\phi^{(')}_c
\eel{e:310}
This is achieved by the choice
\be
&&f_{ab}^{\;\;\;c}\equiv \halv(H^c_{\;\;ab}+H^c_{\;\;ba})\equiv
-g_{ab}^{\;\;\;c}
\eel{e:311}
in \r{309} in which case we get
\be
&&\phi_a=\psi'_a+\halv
H^b_{\;\;ab}-i\pi_bM^b_{\;\;a}-H^b_{\;\;ca}k^{\dag}_bc^c+H^b_{\;\;ca}c^{\dag
c}k_b \nn\\
&&\phi'_a=\phi_a-H^b_{\;\;ab}
\eel{e:312}

\section{Solving the BRST conditions.}
\setcounter{equation}{0}
	By means of a bigrading generated by $c^{\dag a}k_a$ and $k_a^{\dag}c^a$ the
BRST condition
$Q|ph\hb=0$ may be replaced by
\be
&&\del|ph\hb=\del^{\dag}|ph\hb=0
\eel{e:401}
The form \r{218} of $\del$ implies then that these conditions are most
naturally solved by
\be
&&c^a|ph\hb=0,\;\;\;\phi_a|ph\hb=0
\eel{e:402}
or
\be
&&c^{\dag a}|ph\hb=0,\;\;\;{\phi'_a}^{\dag}|ph\hb=0
\eel{e:403}

That all nontrivial solutions of \r{401} are contained in \r{402}-\r{403}
follow \eg from the
general analysis of auxiliary conditions in \cite{Aux}. If one imposes
\be
&&k_a|ph\hb=0
\eel{e:4031}
then consistency requires
\be
&&[Q, k_a]|ph\hb=\phi_a|ph\hb=0
\eel{e:4032}
The solutions of these conditions contain zero norm states, $|ph\hb\sim
k^{\dag}_a\cdots|\;\;\hb$
which are killed by the condition $c^a|ph\hb=0$. Hence, we have arrived at
\r{402}. Similarly one
may start with
\be
&&k^{\dag}_a|ph\hb=0
\eel{e:4033}
which requires
\be
&&[Q, k^{\dag}_a]|ph\hb={\phi'_a}^{\dag}|ph\hb=0
\eel{e:4034}
for consistency. The solutions contain also here zero norm states if one does
not impose $c^{\dag
a}|ph\hb=0$ which makes us arrive at \r{403}. Now, \r{401} has always solutions
satisfying
\be
&&c^a|ph\hb=c^{\dag
a}|ph\hb=0
\eel{e:4035}
which are zero norm states. These states may however couple to solutions of
\r{401} satisfying
\be
&&k_a|ph\hb=k^{\dag
}_a|ph\hb=0
\eel{e:4036}
which requires
\be
&&\phi_a|ph\hb=(\phi'_a)^{\dag}|ph\hb=0
\eel{e:4037}
which are equivalent to
\be
&&\phi_a|ph\hb=0,\;\;\;M^b_{\;\;a}\pi_b|ph\hb=0
\eel{e:4038}
The last conditions requires
\be
&&\pi_a|ph\hb=0
\eel{e:4039}
for some $a$ which does not allow for any solution belonging to an inner
product space. Hence, there
are no solutions of  \r{4036}. In conclusion, the nontrivial solutions of
\r{402} or \r{403} contain the nontrivial solutions of \r{401}. (A similar
argument also applies for
the treatment of \cite{Simple}.)

In order to solve the equations \r{402} and \r{403} we first introduce the
transformed variables (cf.
\cite{Simple}) \be
&&\tilde{\pet_a}\equiv e^A\pet_ae^{-A},\;\;\;\tilde{\bapet_a}\equiv
e^A\bapet_ae^{-A},
\;\;\;\tilde{\psi_a}\equiv e^A\psi_ae^{-A}
\eel{e:404}
where $A$ is the hermitian operator
\be
&&A\equiv[\rho, Q],\;\;\;\rho\equiv\bata_a\chi^a
\eel{e:405}
It is explicitly
\be
&&A=\pi_a\chi^a-i\bata_a\eta^bM^a_{\;\;b}
\eel{e:406}
If we in addition to the previous assumption that all components of
$M^a_{\;\;b}$ and $[M^a_{\;\;b},
\psi_c]$ commute also assume that
\be
&&[\chi^a, M^b_{\;\;c}]=0
\eel{e:407}
then we get the following relations from the definitions \r{404}
\be
&&\tilde{\pet_a}=k_a,\;\;\;\tilde{\bapet_a}=2iM^a_{\;\;b}c^b,\;\;\;\phi_a=
\tilde{\psi_a}+\halv iU_{ab}^{\;\;\;b}  +(c^{\dag c}k_b-k_bc^c)H^b_{\;\;ca}
\eel{e:408}
and
\be
&&{\tilde{\pet_a}}^{\dag}=k^{\dag}_a,\;\;\;{\tilde{\bapet_a}}^{\dag}=-2iM^a_{\;\;b}c^{\dag
b},\nn\\
&&{\phi'_a}^{\dag}={\tilde{\psi_a}}^{\dag}+\halv iU_{ab}^{\;\;\;b}
+(c^ck^{\dag}_b-k^{\dag}_bc^{\dag c})H^b_{\;\;ca}
\eel{e:409}
Since the solutions of \r{402} which do not satisfy
\be
&&k_a|ph\hb=0
\eel{e:410}
have zero norms, we find from \r{408} that the non-trivial solutions of \r{402}
have the form
\be
&&|ph\hb=e^A|\phi\hb
\eel{e:411}
where $|\phi\hb$ satisfies
\be
&&(\psi_a+\halv i
U_{ab}^{\;\;\;b})|\phi\hb=0,\;\;\;\pet_a|\phi\hb=\bapet^a|\phi\hb=0
\eel{e:412}
Notice that this implies that \r{411} has ghost number zero.
The non-trivial solutions of \r{403} are on the other hand of the form
\be
&&|ph\hb=e^{-A}|\phi\hb
\eel{e:413}
where $|\phi\hb$ satisfies \r{412}. The first condition in \r{412} may be
replaced by
\be
&&[Q, \pet_a]|\phi\hb=0
\eel{e:414}
Since
\be
&&[Q, \pet_a]=\psi_a+\psi_a^{gh},\;\;\;\psi_a^{gh}\equiv \halv i
 U_{ab}^{\;\;\;c}(\pet_c\eta^b-\eta^b\pet_c)
\eel{e:415}
eqn \r{414} reduces to the first condition in \r{412} when the last conditions
in \r{412} are
applied.

In distinction to the case in \cite{Simple} conditions \r{412} are not
trivially solved. Instead of
an explicit solution we have obtained a relation between Dirac quantization and
BRST quantization.

\section{Gauge fixing and abelianization.}
\setcounter{equation}{0}
The unitary transformation \r{302}-\r{303} only transformed the antighosts and
not the ghosts. We
may also consider a unitary transformation which instead transforms the ghosts
and not the
antighosts. We consider therefore
\be
&&\eta'^a=M^a_{\;\;b}\eta^b,
\;\;\;\pet'_a=(M^{-1})^b_{\;\;a}\pet_b,\;\;\;\bata'_a=\bata_a
,\;\;\;\bapet'^a=\bapet^a
\eel{e:501}
This transformation leads also to a decomposition \r{1} but with  a different
expression for $\del$. To see this we first define a gauge generator which
commutes with the new
ghosts. We find
\be
&&\psi''_a\equiv\psi_a+\halv(\eta^c\pet_b-\pet_b\eta^c)H^b_{\;\;ca}\nn\\
&&=\psi_a+\halv(\eta'^c\pet'_b-\pet'_b\eta'^c)G^b_{\;\;ca}
\eel{e:502}
where  $G^b_{\;\;ca}$ and $H^b_{\;\;ca}$ are defined in \r{305}. As before we
assume that all
components of $M^c_{\;\;e}$ and $[M^b_{\;\;d}, \psi_a]$ commute. One may then
show that
\be
&&[\psi''_a, \psi''_b]=iU_{ab}^{\;\;c}\psi''_c
\eel{e:503}
It is now straight-forward to show that the BRST charge \r{201} may be written
as \r{1} where
\be
&&\del=c^{\dag a}A_a
\eel{e:504}
and where
\be
&&A_a\equiv(M^{-1})^b_{\;\;a}\psi'''_b=
\psi''''_b(M^{-1})^b_{\;\;a}
\eel{e:505}
where in turn
\be
&&\psi'''_a\equiv\psi''_a+\halv
H^b_{\;\;ab}-i\pi_cM^c_{\;\;a},\;\;\;\psi''''_a\equiv\psi'''_a-H^b_{\;\;ab}
\eel{e:506}
which both satisfy the algebra \r{503}. One may furthermore show that
\be
&&[A_a, A_b]=0,
\eel{e:507}
which means that we have obtained a BRST charge with an abelian gauge symmetry.
Similar
abelianizations have been considered in many papers before \cite{BF,MH,SHA}.
However, such a compact
general expression has not been given although the corresponding classical
expressions
may be extracted from \cite{SHA}.
The conditions $\del|ph\hb=\del^{\dag}|ph\hb=0$ are here naturally solved by
\be
&&c^a|ph\hb=0,\;\;\;A_a|ph\hb=0
\eel{e:508}
or
\be
&&c^{\dag a}|ph\hb=0,\;\;\;{A_a}^{\dag}|ph\hb=0
\eel{e:509}
which are equivalent to
\be
&&c^a|ph\hb=0,\;\;\;\psi'''_a|ph\hb=0
\eel{e:510}
or
\be
&&c^{\dag a}|ph\hb=0,\;\;\;{\psi''''_a}^{\dag}|ph\hb=0
\eel{e:511}
respectively. (As in section 3 these conditions contain all nontrivial
solutions of
$\del|ph\hb=\del^{\dag}|ph\hb=0$.)
For the operators defined in \r{404} we find here under the same
assumptions as in the previous section \be
&&\tilde{\pet_a}=M^b_{\;\;a}k_b,\;\;\;\tilde{\bapet^a}=2ic^a\;\;\;\phi'''_a=
\tilde{\psi_a}+\halv iU_{ab}^{\;\;\;b}  +(c^{\dag
c}k_b-k_bc^c)G^b_{\;\;ca},\nn\\
&&{\psi''''_a}^{\dag}={\tilde{\psi_a}}^{\dag}+\halv iU_{ab}^{\;\;\;b}
+(c^ck^{\dag}_b-k^{\dag}_bc^{\dag c})G^b_{\;\;ca}
\eel{e:512}
Hence, \r{509}, \r{511} has the solution \r{411}, and \r{510}, \r{512} the
solution \r{413}.

\section{Generalizations}
\setcounter{equation}{0}
In the above construction the solutions  \r{411} and \r{413} were only obtained
under the
restriction \r{407}, \ie
\be
&[\chi^a, M^b_{\;\;c}]=0
\eel{e:601}
This restriction may be relaxed if  the unitary transformations \r{302},\r{303}
and \r{501} are
replace  by
\be
&&\bata'_a=M'^b_{\;\;a}\bata_b,\;\;\;
\bapet'^a=(M'^{-1})^a_{\;\;b}(\bapet^b-\Delta M^b_{\;\;c}\eta^c)\nn\\
&&\eta'^a=\eta^a,\;\;\;\pet'_a=\pet_a+\Delta M^b_{\;\;a}\bata_b
\eel{e:602}
and
\be
&&\eta'^a=M'^a_{\;\;b}\eta^b,
\;\;\;\pet'_a=(M'^{-1})^b_{\;\;a}(\pet_b+\Delta
M^c_{\;\;b}\bata_c)\nn\\&&\bata'_a=\bata_a,\;\;\;\bapet'^a=\bapet^a-\Delta
M^a_{\;\;b}\eta^b
\eel{e:603}
respectively, where
\be
&&M'^a_{\;\;b}\equiv
M^a_{\;\;b}+\sum_{n=1}^{\infty}\frac{1}{(2n+1)!}[\chi^{c_1},
[\chi^{c_2},\cdots[\chi^{c_{2n}},M^a_{\;\;b}]\pi_{c_1}\pi_{c_2}\cdots\pi_{c_{2n}}\nn\\
&&\Delta M^a_{\;\;b}\equiv\sum_{n=1}^{\infty}\frac{1}{(2n)!}[\chi^{c_1},
[\chi^{c_2},\cdots[\chi^{c_{2n-1}},M^a_{\;\;b}]\pi_{c_1}\pi_{c_2}\cdots\pi_{c_{2n-1}}
\eel{e:604}
That the components of
$M'^a_{\;\;b}$ commute follows from the commutability of $M^a_{\;\;b}$ under
rather weak conditions.
In particular if we have
\be
&&[\chi^{c_1}, [\chi^{c_2}, M^a_{\;\;b}]]\pi_{c_1}\pi_{c_2}=0
\eel{e:605} then no further condition is necessary.

The example in section 9 shows that there exist
even further generalizations. We may \eg relax commutability of the matrices.

\section{Further properties of the general solutions.}
\setcounter{equation}{0}
We have found unitary transformations leading to a natural bigrading such that
$\del|ph\hb=\del^{\dag}|ph\hb=0$ has the solutions
\be
&&|ph\hb_{\pm}=e^{\pm[\rho, Q]}|\phi\hb
\eel{e:701}
where $\rho$ is given by \r{405} and where $|\phi\hb$ satisfies \r{412}.  In
fact, we have obtained
a whole class of solutions since there is a large freedom in the choice of
gauge fixing variable
$\chi^a$ in $\rho$. We expect that the properties of the physical state space
in general does not
depend on the chosen bigrading. We should therefore have
\be
&&_{\chi}\vb ph|ph\hb_{\chi}\propto\,_{\chi'}\vb ph|ph\hb_{\chi'}
\eel{e:702}
where $|ph\hb_{\chi}$ is one of the solutions \r{701} with gauge fixing
variable $\chi^a$.
In particular, we may choose $\chi'^a=\al\chi^a$ for any positive real constant
$\al$ in which case
\r{702} looks like the relations given in \cite{Simple}. We expect to have
equality in \r{702} when
$|ph\hb_{\chi}$ and $|ph\hb_{\chi'}$ have the same $|\phi\hb$ state.

Introducing a complete set of states $|x\hb$ in the matter space, \ie
$\int\!dx|x\hb\vb x|=\bett$,
we may obtain the norm
\be
&&_{\chi}\vb ph|ph\hb_{\chi}\propto\int\!dx\del^m(\chi(x))\det M(x)|\phi(x)|^2
\eel{e:703}
where $\phi(x)$ are solutions of a Dirac quantization. This representation
follows from the rule
that Lagrange multipliers must be quantized with opposite metric states to the
gauge fixing variable
$\chi^a$. Choosing positive metric states for the matter space requires
indefinite ones for the
Lagrange multipliers which implies that $\pi_a$ has imaginary eigenvalues in
\r{703} \cite{Gen}.

\section{Example 1: The relativistic particle.}
\setcounter{equation}{0}
Consider the manifestly Lorentz covariant and hermitian coordinate and momentum
operators
$X^{\mu}$ and $P^{\mu}$ satisfying
 \be
&&[X^{\mu}, P^{\nu}]_-=i\eta^{\mu\nu}
\eel{e:801}
where $\eta^{\mu\nu}$ is a space-like Minkowski metric, \ie diag
$\eta^{\mu\nu}=(-1,+1,+1,+1)$.
 A free massless relativistic particle is described by the
 BRST charge operator
\be
&&Q=\halv(P^2+m^2)\eta+\pi\bapet
\eel{e:802}
where $P^2+m^2$ is the mass shell operator. A typical gauge fixing operator is
\be
&&\chi\equiv X^0-\tau
\eel{e:803}
It gives rise to the matrix operator
\be
&&[\chi, \halv(P^2+m^2)]=iP^0
\eel{e:804}
The gauge choice \r{803} requires therefore that we work on states for which
$P^0$ is non-zero. The
analysis of section 3 yields now the decomposition \r{1} where
\be
&&\del=c^{\dag}\phi
\eel{e:805}
where in turn
\be
&&c\equiv\halv(\eta+i\frac{1}{P^0}(\bapet-\halv\pi\eta))\nn\\
&&\phi\equiv\halv(P^2+m^2+\pi^2)+iP^0\pi=\halv({\vec{P}}^2+m^2-(P^0-i\pi)^2)
\eel{e:806}
(In this case the generalized formulas of section 6 must be used.) The
treatment in section 5 leads
on the other hand to \r{803} with the expressions \r{804} replaced by $c\ra P^0
c,\;\phi\ra
\phi/P^0$.

The genuine physical states are according to section 7
\be
&&|ph\hb_{\pm}=e^{\pm[\rho, Q]}|\phi\hb
\eel{e:807}
where
\be
&&\rho=(X^0-\tau)\bata
\eel{e:808}
and where $|\phi\hb$ satisfies
\be
&&(P^2+m^2)|\phi\hb=0\nn\\
&&\pet|\phi\hb=\bapet|\phi\hb=0
\eel{e:809}
The norms are
\be
&&_{\pm}\vb ph|ph\hb_{\pm}=\vb\phi|e^{\pm2[\rho,Q]}|\phi\hb=\vb\phi|
e^{\mp 2iP^0\bata\eta}e^{\pm2\pi(X^0-\tau)}e^{\mp
2iP^0\bata\eta}|\phi\hb=\nn\\&&=\pm Ci\int d^4x
d\pi(\phi_0^*(x,\pi)e^{i\pi(x^0-\tau)}{\partial_0}\phi_0(x,\pi)-{\partial_0}
\phi_0^*(x,\pi)e^{i\pi(x^0-\tau)}\phi_0(x,\pi))=\nn\\&&=\pm Ci\int
d^3x\phi_0^*(x,0)\stackrel{\lra}{\partial_0}\phi_0(x,0)
\eel{e:810}
where
\be
&&C\equiv_{\pet\bapet}\vb 0|i\eta\bata|0\hb_{\pet\bapet}
\eel{e:811}
which is a real constant. It may be chosen to be positive or negative (a choice
of the ghost state
representation). The bosonic vacua are assumed to have positive norms (see
\cite{Propa}). In the
second to last equality we have used the rule that $X^0, P^0$ and $\pi, v$ must
be quantized with
opposite metric states \cite{Gen,Simple}. Notice that $X^0, P^0$ must be
quantized with positive
metric states in order for the BRST condition to yield non-trivial solutions.
(See conditions
\r{402},\r{403} with \r{806} or \r{809}.)  The last equality in \r{810} follows
from the fact that
$\phi_0(x,\pi)$ is a solution of the Klein-Gordon equation.

Obviously \r{810} is not positive definite. It can only be made positive if
$P^0$ is restricted to
be positive or negative in the original state space. However, this is certainly
not a covariant
choice in the original state space. This means that we have not obtained a
completely consistent
quantization of the spinless particle. (In \cite{Propa} a different consistent
quantization is given.)

\section{Example 2: The spinning relativistic particle.}
\setcounter{equation}{0}

The worldline supersymmetric particle model \cite{Halv} involves apart from the
coordinate and
momentum operators $X^{\mu}$ and $P^{\mu}$ also the hermitian fermionic
four-vector $\ga^{\mu}$
satisfying \be
&&[\ga^{\mu}, \ga^{\nu}]_+=-2\eta^{\mu\nu}
\eel{e:901}
where $\eta^{\mu\nu}$ is the space-like Minkowski metric. (In
the matrix representation $\ga^{\mu}$ are the ordinary Dirac gamma matrices.)
For the description of
a massless spin $1/2$ particle this is all we need. The constraint variables
are here $P^2$ and
$P\cdot\ga$ satisfying the world-line supersymmetry algebra. The BRST charge
operator is from
\cite{BV}
 \be
 &&Q=P^2\eta+P\cdot\ga
c+\pet c^2+\pi\bapet+\kappa\bak \eel{e:902}
where the variables satisfy the following (anti-)commutation relations apart
from \r{901} (the nonzero
part):
 \be
&&[X^{\mu}, P^{\nu}]_-=i\eta^{\mu\nu},
\;\;\;[\pi, v]_-=-i,\;\;\;[\kappa, \la]_+=1,\nn\\
&&[\pet, \eta]_+=1,\;\;\;[\bapet, \bata]_+=1,\;\;\;[k, c]_-=-i,\;\;\;[\bak,
\bac]_-=-i
\eel{e:903}
where $k,c$ are bosonic ghosts and $\bak, \bac$ the corresponding antighosts,
$\la$ is a fermionic
Lagrange multiplier and $\kappa$ its conjugate momentum.

The above model is a model with a graded symmetry involving \eg bosonic ghosts.
The previous general
formulas do not directly apply here. However, since we expect that the general
treatment may be
straight-forwardly extended to models with graded gauge symmetries we have
partly chosen the
spinning particle model to demonstrate the viability of this expectation.

	A natural choice of gauge variables is
\be
&&\chi^1\equiv X^0-\tau,\;\;\;\chi^2\equiv\ga^0
\eel{e:904}
They lead to the following hermitian matrix components ($M^a_{\;\;b}=
(i)[\chi^a, \psi_b]$)
\be
&&M^1_{\;\;1}=2P_0,\;\;\;M^1_{\;\;2}=-\ga^0,\;\;\;  M^2_{\;\;1}= 0,\;\;\;
M^2_{\;\;2}=2P_0
\eel{e:905}
The inverse is
\be
&&(M^{-1})^1_{\;\;1}=\frac{1}{2P_0},\;\;\;(M^{-1})^1_{\;\;2}=\frac{\ga^0}{(2P_0)^2}\nn\\&&
(M^{-1})^2_{\;\;1}= 0,\;\;\;  (M^{-1})^2_{\;\;2}=\frac{1}{2P_0}
 \eel{e:906}
The treatment of sections 3 and 6 is helpful in deriving  the following
appropriate unitary
transformation \be
&&\bapet\; \ra\; \bapet'\equiv\frac{1}{2P_0}(\bapet-\pi\eta+\halv\kappa c+\bata
c^2+\frac{\ga^0}{2P_0}(\bak-\pi c))\nn\\
&&\bak\; \ra\; \bak'\equiv\frac{1}{2P_0}(\bak-\pi c),\;\;\;\eta\; \ra\;
\eta'=\eta,\;\;\;c\; \ra\;
c'=c \eel{e:907}
which leads to the decomposition \r{1} where
\be
&&\del=c^{\dag}_1\phi_1+c^{\dag}_2\phi_2
\eel{e:908}
where in turn
\be
&&c_1\equiv\halv(\eta'-i\bapet'),\;\;\;c_2\equiv\halv(c'-i\bak')\nn\\
&&\phi_1\equiv P^2+\pi^2-2iP_0\pi,\;\;\;\phi_2\equiv P\cdot\ga+\pet c-\pi\bata
c+i(\pi\ga^0-2P_0\kappa)
\eel{e:909}
Notice that $\phi_1$ and $\phi_2$ satisfy the same algebra as $P^2$ and
$P\cdot\ga$ and that
$\phi_1$ is equal to  $\phi$ in \r{806} apart from a factor 2.

If we instead had followed the treatment of section 5 we would obtain another
expression for $\del$
than \r{908}. However, with the choice \r{904} of gauge fixing variables we
would not obtain
abelianization since the matrix component $M^1_{\;\;2}=-\ga^0$ does not
anticommute with itself.
Abelianization is only obtained for a light-cone gauge $\chi^1\equiv
X^+-\tau,\;\chi^2\equiv\ga^+$
in which case the abelian constraints have the structure
$P\cdot\ga+\frac{\ga^+}{2P^+}P^2$ and $P^2$.

The solutions of the conditions \r{402} and \r{403} are according to section 7
\be
&&|ph\hb_{\pm}=e^{\pm[\rho, Q]}|\phi\hb
\eel{e:910}
where
\be
&&\rho\equiv\bata(X^0-\tau)+\bac\ga^0
\eel{e:911}
and where $|\phi\hb$ satisfies
\be
&&\bapet|\phi\hb=\bak|\phi\hb=0,\;\;\;P^2|\phi\hb=(P\cdot\ga+\pet c)|\phi\hb=0
\eel{e:912}
which makes $|\phi\hb$ BRST invariant. Since only the zero ghost sector will
contribute to the norms
we may impose the additional conditions
\be
&&\pet|\phi\hb_0=k|\phi\hb_0=0
\eel{e:913}
Obviously $|\phi\hb_0$ then satisfies  the Dirac equation in the wave function
representation. The
norm of the physical state \r{910} is
\be
&&_{\pm}\vb ph|ph\hb_{\pm}=\,_0\vb\phi|e^{\pm\al[\rho, Q]}|\phi\hb_0
\eel{e:914}
where $\al=2$. Now one may show that the right hand side yields the same value
for any positive real
$\al$. We choose therefore $\al=1$ for simplicity. We have then
\be
&&_0\vb\phi|e^{\pm[\rho,
Q]}|\phi\hb_0=\,_0\vb\phi|e^{\pm(\pi(X^0-\tau)+i\ga^0\kappa+2P_0\bac
c-2iP_0\bata\eta)}|\phi\hb_0=\nn\\ &&=_0\vb\phi|e^{\mp iP_0\bata\eta}e^{\pm
P_0\bac
c}e^{\pm\pi(X^0-\tau)}e^{\pm P_0\bac
c}e^{\mp iP_0\bata\eta}e^{\pm i\ga^0\kappa}|\phi\hb_0=\nn\\ &&=-C\int
d^3x(\bar{\phi_0}\ga^0\ga^5\phi_0|_+-\bar{\phi_0}\ga^0\ga^5\phi_0|_-)
\eel{e:915}
where $C$ is given by \r{811} and where $\phi_0(x)$ is a solution of the Dirac
equation and where
$|_{\pm}$ denotes the restriction to positive/negative energy solutions.
(Bosonic vacua are ssumed
to have positive norms.)   In order to obtain the last equality in \r{915} one
has to use two rules
\cite{Propa}: (i) $X^0, P^0$ and $\pi, v$ are quantized with opposite metric
states, and (ii) bosonic
ghosts and antighosts must be quantized with opposite metric states. (Here
$X^0, P^0$ must be
quantized with positive metric states otherwise there exist no nontrivial
solutions at all.)
Eq.\r{915} should be compared with eq. (4.23) in \cite{MO}.

Since the two chiralities as
well as the positive and negative energy solutions have opposite norms we have
not obtained a
consistent quantization of the spin $1/2$ model without further projections.
(Another treatment is
given in \cite{Propa}.)

  \end{document}